\documentclass[fleqn,usenatbib]{mnras}

\usepackage{mathptmx}

\usepackage[T1]{fontenc}
\usepackage{ae,aecompl}

\usepackage{graphicx}	
\usepackage{amsmath}	
\usepackage{amssymb}	

\usepackage[none]{hyphenat}

\def\PMO{$^{1}$}
\def\CAS{$^{2}$}
\def\Curtin{$^{3}$}
\def\CAASTRO{$^{4}$}
\def\geo{$^{5}$}
\def\ASU{$^{6}$}

\def\USydney{$^{7}$}
\def\UToronto{$^{8}$}
\def\Victoria{$^{9}$}
\def\UWisc{$^{10}$}
\def\UW{$^{11}$}
\def\UWA{$^{12}$}
\def\CAASTROD{$^{13}$}

\title[Radio emission from meteors]{Limits on radio emission from meteors using the MWA}

\author[X. Zhang et al.]{
	Xiang Zhang\PMO$^,$\CAS$^, $\Curtin$^, $\CAASTRO\thanks{E-mail: zhangxiang@pmo.ac.cn}, 
	Paul Hancock\Curtin$^,$\CAASTRO, 
	Hadrien A. R. Devillepoix\geo, 
	Randall B. Wayth\Curtin$^,$\CAASTRO, 
	\newauthor
	A.~Beardsley\ASU, 
	B.~Crosse\Curtin, 
	D.~Emrich\Curtin, 
	T.~M.~O.~Franzen\Curtin, 
	B.~M.~Gaensler\USydney$^,$\CAASTRO$^,$\UToronto, 
	\newauthor
	L.~Horsley\Curtin, 
	M.~Johnston-Hollitt\Curtin$^,$\Victoria, 
	D.~L.~Kaplan\UWisc, 
	D.~Kenney\Curtin,
	M.~F.~Morales\UW, 
	\newauthor
	D.~Pallot\UWA,
	K.~Steele\Curtin,
	S.~J.~Tingay\Curtin$^,$\CAASTRO,
	C.~M.~Trott\Curtin$^,$\CAASTRO$^,$\CAASTROD, 
	M.~Walker\Curtin,
	A.~Williams\Curtin, 
	\newauthor
	C.~Wu\UWA,
	Jianghui Ji\PMO,
	and Yuehua Ma\PMO.
	\\
	$^{1}$CAS Key Laboratory of Planetary Sciences, Purple Mountain Observatory, Chinese Academy of Sciences, Nanjing 210008, China\\
	$^{2}$University of Chinese Academy of Sciences, Beijing 100049, China\\
	$^{3}$International Centre for Radio Astronomy Research, Curtin University, Bentley, WA 6102, Australia\\
	$^{4}$ARC Centre of Excellence for All-Sky Astrophysics (CAASTRO), Sydney, NSW 2006, Australia\\
	$^{5}$School of Earth and Planetary Sciences, Curtin University, Bentley, WA 6102, Australia\\
	$^{6}$School of Earth and Space Exploration, Arizona State University, Tempe, AZ 85287, USA\\	
	$^{7}$Sydney Institute for Astronomy, School of Physics, The University of Sydney, NSW 2006, Australia\\
	$^{8}$Dunlap Institute for Astronomy and Astrophysics, University of Toronto, ON, M5S 3H4, Canada\\
	$^{9}$Peripety Scientific Ltd., PO Box 11355 Manners Street, Wellington 6142, New Zealand\\
	$^{10}$Department of Physics, University of Wisconsin--Milwaukee,
	Milwaukee, WI 53201, USA\\
	$^{11}$Department of Physics, University of Washington, Seattle, WA 98195, USA\\
	$^{12}$International Centre for Radio Astronomy Research, University of Western Australia, Crawley 6009, Australia\\
	$^{13}$ARC Centre of Excellence for All Sky Astrophysics in 3 Dimensions (ASTRO 3D), Bentley WA, Australia
}

\date{Accepted XXX. Received YYY; in original form ZZZ}

\pubyear{2017}

\begin{document}
\label{firstpage}
\pagerange{\pageref{firstpage}--\pageref{lastpage}}
\maketitle

\begin{abstract}
Recently, low frequency, broadband radio emission has been observed accompanying bright meteors by the Long Wavelength Array (LWA). The broadband spectra between 20 and 60 MHz were captured for several events, while the spectral index (dependence of flux density on frequency, with $S_\nu \propto \nu^\alpha$) was estimated to be $-4\pm1$ during the peak of meteor afterglows. Here we present a survey of meteor emission and other transient events using the Murchison Widefield Array (MWA) at 72-103 MHz. In our 322-hour survey, down to a $5\sigma$ detection threshold of 3.5 Jy/beam, no transient candidates were identified as intrinsic emission from meteors. We derived an upper limit of -3.7 (95\% confidence limit) on the spectral index in our frequency range. We also report detections of other transient events, like reflected FM broadcast signals from small satellites, conclusively demonstrating the ability of the MWA to detect and track space debris on scales as small as 0.1 m in low Earth orbits.
\end{abstract}

\begin{keywords}
plasmas -- meteorites, meteors, meteoroids -- radio continuum: transients
\end{keywords}



\section{Introduction}


When a rocky or metallic object (meteoroid) plunges into the atmosphere and is heated to incandescence, a meteor can be observed.
Very bright meteors are referred to as fireballs. 
The meteor is heated by radiation from the atmospheric shock front that it produces \citep{de2015planetary}, causing iron and silicates to melt and vaporize. 
The vaporized atoms are ionized in collisions with air atoms, producing a cloud of quasi-neutral plasma, which is referred to as the ionized trail \citep{dokuchaev1960formation}. The ionized trails are known to reflect radio waves, and radio echoes are used to measure the orbits of meteors and radiants of meteor showers \citep{ceplecha1998meteor}.

Detailed investigations of radio emission from meteors began in the 1950s, when astronomers proposed that plasma resonance in meteor trails might produce radio noise \citep{hawkins1958search}. 
Detections of low frequency emission were reported to be coincident with large meteors in the past several decades \citep{beech1995vlf, guha2012investigation}. 
Sometimes bright meteors were observed accompanied by acoustic propagation, which might be caused by radio emission converting into electrophonic sounds \citep{keay1980anomalous, keay1992electrophonic, keay1994rate}.

Recently, scientists working with the Long Wavelength Array (LWA) \citep{taylor2012first, ellingson2013design} made some interesting discoveries of radio emission from meteors. 
In 2014, two transient events were reported in a search for prompt low-frequency emission from Gamma-Ray Bursts (GRBs) with the first station of the Long Wavelength Array (LWA1) \citep{obenberger2014limits}. 
The LWA1 was operating in the narrow transient buffer mode, with a usable bandwidth of 75 kHz tunable to any centre frequency between 10 and 88 MHz. 
These two events lasted for 75 and 100s, respectively, at 37.9 and 29.9 MHz. 
They were not coincident with any known GRBs. 
Further observations revealed more similar long-duration (tens of seconds) transients. 
Many of these transients were coincident with optical meteors, both spatially and temporally. 
Between April 2014 and April 2016, a total of ~20,000 hours data were collected, in which 154 radio transients were detected \citep{obenberger2016altitudinal}. 
Optical meteor counterparts were coincident with 44 of these radio transients.


The transients correlated with meteors are different from the well-studied radio echoes from meteor trails \citep{obenberger2014detection}. 
First, most radio transmitters are polarized, thus the reflections from meteor trails are also polarized \citep{close2011polarization, helmboldt2014all}. 
However, no significant amount of polarization has shown up in these transient cases, either linear or circular. 
Second, transmitters often broadcast in very narrow radio bands, and spectral lines are visible in meteor reflection, but spectral features are not found anywhere in the LWA1 transients. 
Third, power profiles of these transients resemble each other, but are quite different from meteor reflections. 
All these differences led \citet{obenberger2014detection} to suggest that meteors emit a previously undiscovered low-frequency, non-thermal pulse.

Broadband measurements were also made with the beamformer mode of the LWA1 in order to obtain the dynamic spectra of the transient events \citep{obenberger2016rates}. 
Three beams were formed and pointed around zenith at azimuths of 60\degr, 180\degr, and 240\degr, all with elevations of 87\degr. 
The field of view for each beam is $\sim 50~\mathrm{deg}^2$. 
Compared with the all-sky imager mode, the beamformer mode's field of view is much smaller, leading to fewer detections. 
The broadband spectra of four events were captured between 22.0 and 55.0 MHz. 
The frequency-dependent flux densities of these events were fit to a power law, and the spectral indices were found to be time variable, with the spectrum steepening over time. 


\citet{obenberger2016rates} also discussed the potential for other observatories to measure meteor spectra, including the Murchision Widefield Array (MWA;  \citealp{tingay2013murchison}) in Australia, the Amsterdam-ASTRON Radio Transients Facility and Analysis Center (AARTFAAC; \citealp{prasad2014real}) based on the Low-Frequency Array (LOFAR;  \citealp{van2013lofar}) in the Netherlands, and two additional LWA stations. 
It was concluded that the MWA, with its exceptionally high sensitivity, had the best opportunity to test the high-frequency predictions of meteor radio afterglows. 
However, there are several factors which might prevent the MWA from detecting radio emission from meteors: the high spatial resolution of the MWA will lead to a drop in peak flux density, thus the meteors might still be undetectable for the MWA; the uncertain and time variable spectral index may cause a lower flux density than predicted. 

Based on the research above, we carried out a 322-hour survey for meteor afterglows with the MWA. 
Our work aims to detect the radio afterglow from ionized meteor trails at higher frequencies and put some limits on the meteor radiation spectra. 

In this paper, we begin in Section \ref{sec:observations} with a description of our observations, both radio and optical. 
The data reduction process is given in Section \ref{sec:data}, including preprocessing, imaging and source finding. A brief description of results is given in Section \ref{sec:results}. In Section \ref{sec:discussion}, we discuss the relation between meteor event rates and flux density, followed by an estimated upper limit on meteor radiation spectra. Section \ref{sec:discussion} also contains some other transient events detected in our survey and a discussion of future work. The conclusion is presented in Section \ref{sec:conclusions}.



\section{observations}
\label{sec:observations}

Both radio and optical observations were carried out in this work. Radio observations were done by the MWA, while optical observations were performed by the Desert Fireball Network (DFN; \citealp{bland2012australian}). The implementation of optical observations allows us to compare radio transient events with optical meteors and investigate possible emission from meteors.

\subsection{Radio observations using the MWA}

\begin{table*}
	\centering
	\caption{Meteor showers observed by the MWA. ZHR stands for Zenithal Hourly Rate, a calculated maximum number of meteors per hour an ideal observer would see in perfectly clear skies with the shower radiant overhead. Velocities listed in this table are apparent meteoric velocities. }
	\label{tab:showers}
	\begin{tabular}{cccccc} 
		\hline
		Date & Name & Total observation length (hour) & Radiant (RA, DEC) & ZHR & Velocity in km/s\\
		\hline
		Dec 14, 2014-2016 &  Geminids &28 & 112\degr, +30\degr & 120 & 35\\
		Feb 08, 2016 &  $\alpha$-Centaurids &2 & 210\degr, -59\degr & 6 & 56\\
		Mar 14, 2016 & $\gamma$-Normids & 2 & 239\degr, -50\degr & 6 & 56\\
		Apr 23, 2016 & $\pi$-Puppids & 2 & 110\degr, -45\degr & Variable & 18\\
		May 05, 2016 & $\eta$-Aquariids & 2 & 338\degr, -01\degr & 40 & 66\\
		\hline
	\end{tabular}
\end{table*}

The MWA is one of the Square Kilometre Array (SKA) Precursor telescopes. It is located at the Murchison Radio-astronomy Observatory in Western Australia, where the Radio Frequency Interference (RFI) is extremely low \citep{offringa2015low}. The MWA consists of 128 aperture array antennas (referred to as tiles) distributed over a $\sim$3-km diameter area. It is optimized for the 80-300 MHz frequency range, with a processed bandwidth of 30.72 MHz for both linear polarizations \citep{tingay2013murchison}.


In this project, radio observations were carried out under two modes, one targeted and one opportunistic. For both observations, all the tiles of the MWA were pointed to the zenith, and the lowest band of the MWA (72.3 - 103.0 MHz) was used. The choice of observational band was based on two reasons: first, previously detected radio emission from meteors was below 60 MHz; second, the chosen band partly overlapped with the FM broadcast band in Australia (87.5 - 108 MHz), making it possible for us to observe reflection and intrinsic emission from meteors at the same time.

Under the targeted mode, we observed several known meteor showers\footnote{\url{https://www.imo.net/}}, which are given in Table \ref{tab:showers}. The $\alpha$-Centaurids, $\gamma$-Normids, $\pi$-Puppids and $\eta$-Aquariids showers were observed because they all have radiants in the Southern Hemisphere. For each of these showers, we observed for two hours around midnight. The Geminids shower, however, was chosen due to its especially high zenithal hourly rate. We observed the Geminids three times -- in 2014, 2015 and 2016. Each time the observation lasted about 9 hours, from dusk to dawn.

For all the meteor showers listed in Table \ref{tab:showers}, a series of 112-second observations were obtained with a temporal resolution of 0.5 second, which is the highest temporal resolution of the MWA. The frequency resolution was 40 kHz. The drift of the sky during each 112-second scan was accounted for during imaging processing by fixing the phase centre for each observation to be at a certain RA/Dec. 

Since we were not able to predict when and where meteors occur, we also performed some opportunistic observations when the MWA was not occupied by other projects. An example of the opportunistic observations is the filling observations carried out in March 2016, when some of the MWA tiles failed to point due to lightning damage. Since these tiles pointed up by default, we pointed all tiles to the zenith and observed the sky with the entire array. The filling observations were carried out with a duration of 232 seconds and a temporal resolution of 2 seconds. A list of the opportunistic observations is given in Table \ref{tab:filling}. The total radio observational time from both dedicated and opportunistic campaigns adds up to approximately 322 hours.

\begin{table}
	\centering
	\caption{Opportunistic observations, performed when the MWA was not occupied by other projects.}
	\label{tab:filling}
	\begin{tabular}{cc} 
		\hline
		Date & Total observation length (hour) \\
		\hline
		Mar 14 - 22, 2015 &  15\\
		Mar 17 - 29, 2016 &  103\\
		Apr 02 - 14, 2016 & 49\\
		May 01 - 03, 2016 & 22\\
		May 10 - Jun 01, 2016 & 93\\
		Sep 08 - 09, 2016 & 4\\
		\hline
	\end{tabular}
\end{table}

\subsection{Optical observations using the DFN}

\begin{figure}
	\includegraphics[width=\columnwidth]{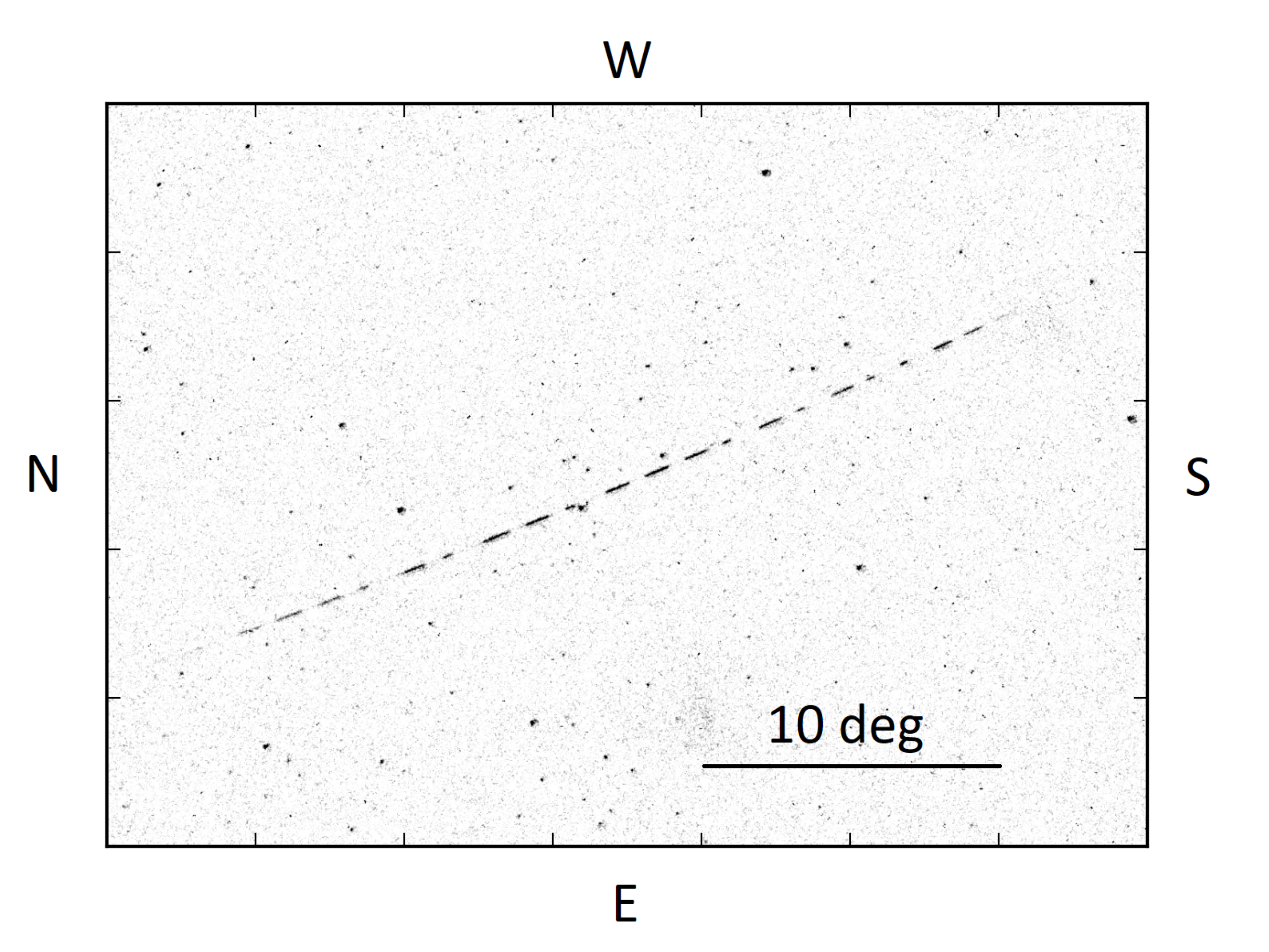}
    \caption{Optical image of a meteor captured by the DFN camera at Wooleen station, on 14 December 2015. The exposure duration was 25 seconds. The gaps in the meteor trail were caused by the coded shutter to measure its apparent speed. }
    \label{fig:optical}
\end{figure}

The DFN is a camera network with dozens of cameras in Western Australia and South Australia. It is designed to detect meteors and triangulate their trajectories, in order to recover the meteorite (debris of a meteor, which impacts with the Earth's surface) and trace the meteor back to its origin. Since September 2015, some DFN cameras have been installed at Wooleen station, 70 km apart from the MWA, thus they share a significant portion of the sky with the MWA. For our 322-hour radio observations, 297 hours were also covered by optical observations.

The Wooleen DFN node includes a standard meteor camera, as well as a specially designed camera for astronomical work \citep{howie2017build}. Both cameras are pointed to the zenith. The standard meteor camera has a fish-eye lens that can see the entire sky. The limiting magnitude for a meteor to be detected by the camera is about 0.5 magnitudes. A modulated liquid crystal shutter is used to determine the angular speed of meteors, which leads to gaps in meteor trails in the images (see Figure \ref{fig:optical}). The standard meteor camera is operated autonomously, taking images with a 29-second exposure time when the Sun is down, and the sky is clear. The astronomical camera, on the other hand, has a much improved sensitivity with a smaller field of view ({80\degr $\times$100\degr}). The exposure duration of the astronomical camera is 13 seconds.

\section{Data reduction}
\label{sec:data}
For radio observations, the data reduction pipeline is composed of four steps: preprocessing via Cotter \citep{offringa2015low}, calibration using bright radio sources, imaging with WSClean \citep{offringa2014wsclean}, and source finding via Aegean \citep{hancock2012compact, 2018PASA...35...11H}. However, in our 322-hour observation, not all the data were of good quality. Therefore, only 308 hours of observational data were processed with the pipeline and used for the analysis.

The optical data were captured as an independent verification of the presence of a meteor. These images were used in their original form (colour JPEGs). Recently a calibration scheme has been created to correct astrometry and photometry of these images, but such calibration was not required in this project.

\subsection{Preprocessing and calibration}
We preprocessed the raw visibility data through the MWA preprocessing pipeline, Cotter, to average the data and convert it into Common Astronomy Software Applications  \citep[CASA]{mcmullin2007casa} measurement set format. Cotter can also flag radio-frequency interference (RFI) with a C++ library provided by the RFI detector, AOFLAGGER \citep{offringa2010post, offringa2012morphological}. 

For each observation, we made two measurement sets with 8-second integration: we flagged RFI in one measurement set (referred to as \textit{emission} data), and kept RFI in another (referred to as \textit{reflection} data). Since ionized meteor trails are known to reflect RFI, the emission data can show intrinsic emission from meteors, while the reflection data are able to reveal reflected radio signal from meteors. However, the RFI flagging process is not able to exclude all the RFI in the FM band, so we only used emission data outside the FM band to make emission images, as described in Section  \ref{sec:imaging}.


After preprocessing, the measurement sets were calibrated using bright point sources with well-modeled emission for the MWA. Based on the models of the calibrators, we derived time-independent, frequency dependent phase and amplitude calibration solutions, which were applied to the measurement sets. 

Calibrators were observed for 112 seconds at the phase centre of the telescope, before or after our scheduled observations for one night. The quality of drift scan data is always limited by the ionosphere, so in calm conditions a single calibration solution can be applied to all the data from the same night \citep{hurley2014murchison}. The most commonly used calibrator was Hydra A. For several nights without useful calibrator observations, we either used a common observation in which a bright radio source was close to the zenith, or used catalogued radio resources from the galactic and extragalactic all-sky MWA survey (GLEAM) \citep{wayth2015gleam, 2017MNRAS.464.1146H} for calibration. 

\subsection{Imaging}
\label{sec:imaging}
The calibrated measurement sets were imaged and deconvolved using WSClean \citep{offringa2014wsclean}, a fast wide-field imager for radio astronomy. For each observation, the reflection and emission measurement sets were separately made into 8-second integrated \textit{reflection} and \textit{emission} all-sky images.
In this process, only short baselines were used, and only frequency channels below the FM band were used for the emission images. Precise limits on baselines and frequency channels were decided via a few tests with a simulated meteor, which are described in Section \ref{sec:simulation}.
The detailed settings of WSClean are listed in Table \ref{tab:wsclean}. 

\begin{table}
	\centering
	\caption{The WSClean settings used to image the meteor observations. All other settings were set to default. }
	\label{tab:wsclean}
	\begin{tabular}{cc} 
		\hline
		Setting  & Value \\
		\hline
		UV range ($\lambda$) & <32\\
        Channel range for emission images (MHz) & 72.3-86.4\\
        Channel range for reflection images (MHz) & 72.3-103.0\\
        Image integration time (s) & 8\\
		Maximum number of clean iterations & 4000\\
		Size of image (pixel) & 240\\
        Size of one pixel (arcmin) & 20\\
        Briggs weighting & 0.5\\
        Polarization & XX, YY\\
		\hline
	\end{tabular}
\end{table}

The selection of image integration time (8 seconds) was based on the detected meteor dynamic spectra from \citet{obenberger2015dynamic}. Meteor emission can last for more than one minute at 20-40 MHz, but the spectral index drops significantly after the first 20 seconds. Thus meteor emission at MWA frequencies may be shorter in duration.

\begin{figure*}
	\includegraphics[width=\textwidth]{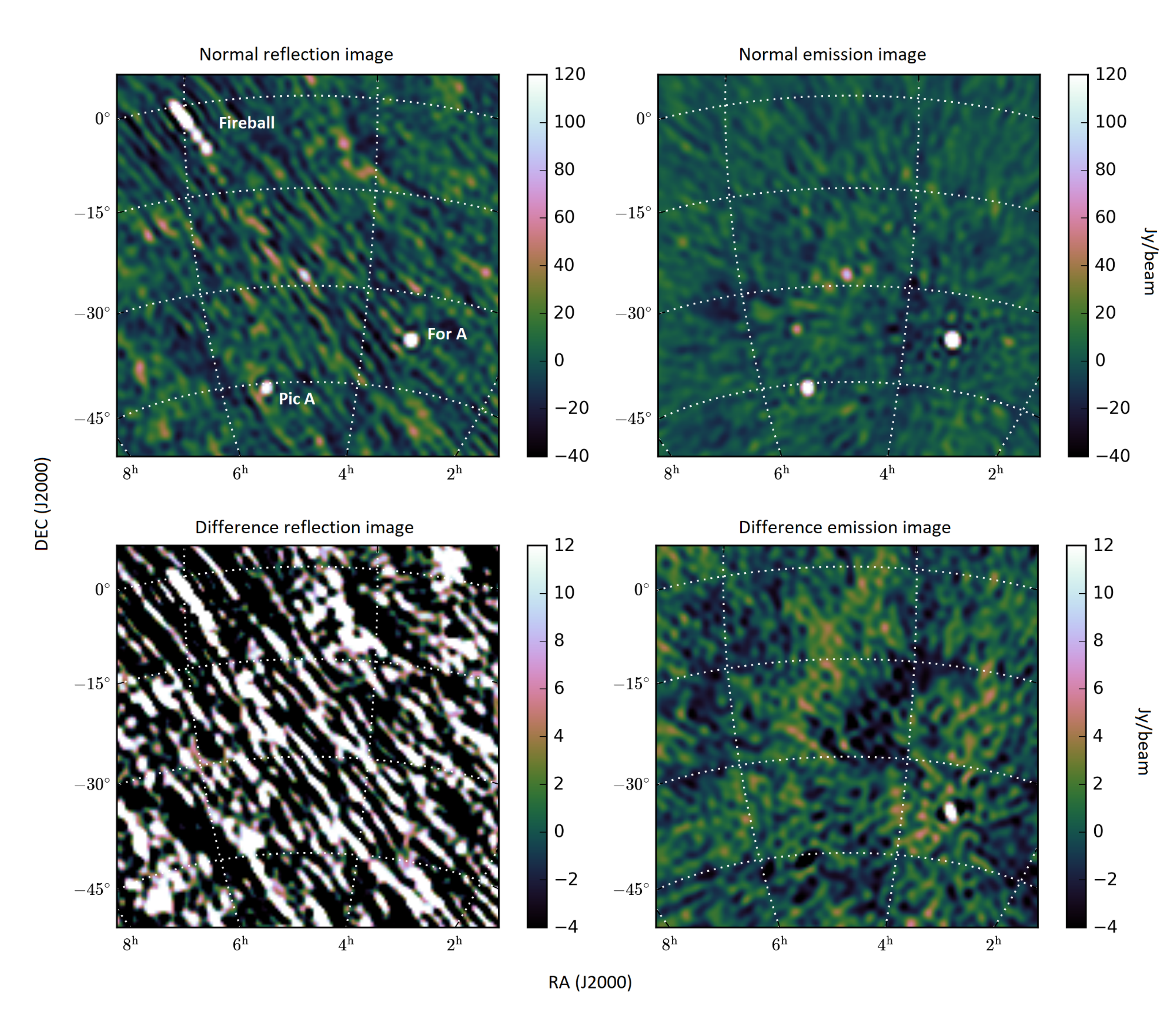}
	\caption{MWA radio images of an 8-second duration, during which a meteor occurred.  Top-left: normal reflection image, showing the radio sky with RFI. Background radio sources like Fornax A and Pictor A can be seen in this image, together with reflected RFI from the ionized meteor trail. Top-right: normal emission image, showing the sky without RFI. This image contains background sources but not reflection from the meteor. Bottom-left: difference reflection image, illustrating temporal variation in the normal reflection image. Bottom-right: difference emission image, showing temporal variation in the normal emission image. This image has capability to reveal intrinsic emission from meteors, but no emission is found for the particular meteor event. Pictor A is removed from this image, while Fornax A remains due to scintillation or instrumental effects. Normal images (top) and difference images (bottom) do not share the same color scale. Compared with \textit{reflection} images (left), \textit{emission} images (right) are based on observational data below FM band, with RFI flagged.}
	\label{fig:frame4}
\end{figure*}

In order to exclude background sources and reduce noise, we subtracted adjacent 8-second snapshots to make difference images, as shown in Figure \ref{fig:frame4}. Dirty (un-deconvolved) images were used for subtraction, because CLEANing may create artifacts in difference images. In the imaging process, the phase centre of each observation was fixed in RA/Dec, so sky rotation didn't introduce artifacts into the difference images.

\subsection{Baseline and channel limits in imaging process}
\label{sec:simulation}

There are two reasons why we only used short baselines. First, like most radio telescopes, the MWA is focused in the far field, since most objects included in the MWA science are effectively at an infinite distance from the telescope. The MWA correlator assumes incoming waves from these sources as plane waves. However, for objects close to the telescope (in the near field), the incoming waves are spherical rather than planar. In antenna design, the widely accepted transition between near field and far field is the Fraunhofer distance $d = 2D^2/\lambda$, where $D$ is the diameter of the telescope and $\lambda$ is the wavelength. For the MWA with its longest baselines at 80 MHz, the Fraunhofer distance is 4800 km. Since the typical height of meteors is 80-120 km, using the longest baselines will put meteors in the near field. In order to place the meteors in the far field at 80 km, we can only use baselines shorter than 387 m. 

Another reason to use the short baselines is to improve the detectability for meteors. A typical meteor trail is tens of kilometres long, several metres wide, and about 100 km above the ground. In MWA observational images, that corresponds to an extended source which is tens of degrees in length and less than one arcmin in width, so we excluded baselines longer than 120 m to get a lower spatial resolution (2.15\degr) and higher peak flux density for meteors. Since the MWA has a 100-m diameter dense core containing 50 tiles \citep{tingay2013murchison}, excluding these baselines increases the thermal noise by just 16\%.

We limited the channel ranges (72.3-86.4 MHz) for emission images mostly due to RFI contamination, since our observational band (72.3-103 MHz) overlapped with the FM broadcasting band in Australia (87.5-108 MHz). Although AOFLAGGER was used to flag RFI, a small amount of RFI was still left within the FM band. Using the channels below the FM band also brought another advantage: the meteor emission is much brighter at lower frequencies, thus only using the lower band can improve the detectability of meteors.

In order to determine the precise baseline and channel limits to be used in the imaging process, we did tests by adding a simulated meteor to the visibility data of a blank sky observation, and made difference emission images with several baseline and channel range settings. 

According to the four meteor events described by \citet{obenberger2016rates}, the spectrum of meteors follows a power law between 20 and 60 MHz. 
The power law is given by $S \propto \nu^\alpha$, where $S$ is the flux density, $\nu$ is the frequency, and $\alpha$ is the spectral index. 
During the peak of the afterglows, $\alpha \sim -4 \pm 1$. 

Here we used one of the four meteors given in \citet{obenberger2016rates} for extrapolation. This meteor belongs to a group of faint/common meteors occurring 130 times  $year^{-1}~\pi~sr^{-1}$. It was detected as an unresolved source by the LWA1 at 25.6 MHz, with a flux density of 1800 Jy. Since the MWA has a higher spatial resolution than the LWA1, and the width of a meteor trail is much smaller than the MWA spatial resolution, we assumed that the simulated meteor was extended in one dimension in MWA images, along the trail. 

\begin{figure}
	\includegraphics[width=\columnwidth]{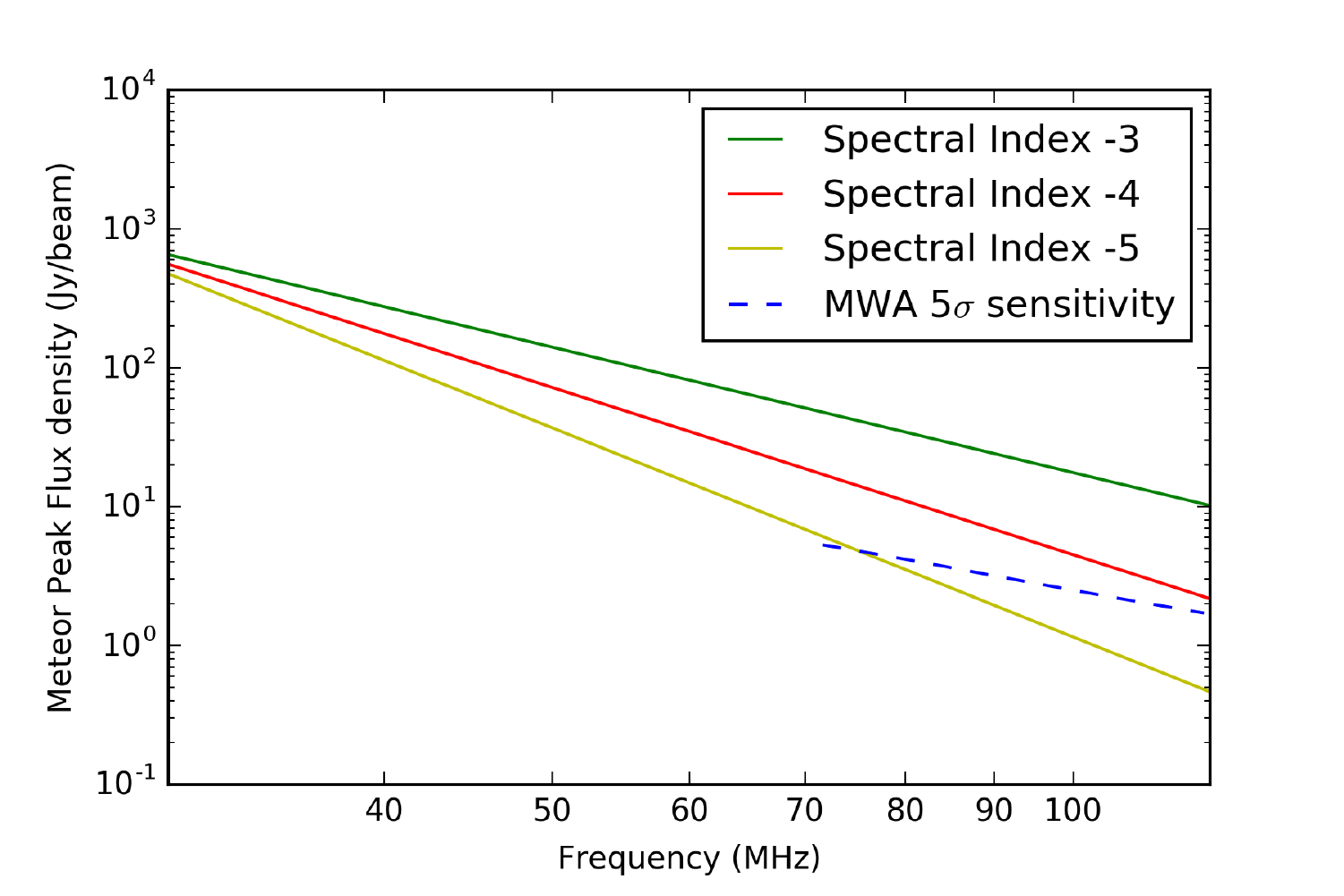}
	\caption{Estimated peak flux densities of the faint/common meteor as observed by the MWA. The solid lines illustrate broadband spectra of meteors with different spectral indices, while the blue dashed line describes the $5\sigma$ sensitivity of the MWA in our observations.}
	\label{fig:radiation}
\end{figure}

Figure \ref{fig:radiation} shows the extrapolated peak flux densities of the meteor, assuming different spectral indices. Also shown are comparison plots of the 5$\sigma$ sensitivity of the MWA (short baselines, 8 second integration and 12 MHz bandwidth). The MWA values were obtained from MWA observational images and data from \citet{sutinjo2015understanding}. As illustrated in Figure \ref{fig:radiation}, it is estimated that in 72-103 MHz band, the MWA is capable of detecting radio emission from meteors when $\alpha >= -4$, but it is not able to detect meteors when $\alpha <=-5$.

\begin{figure}
	\includegraphics[width=\columnwidth]{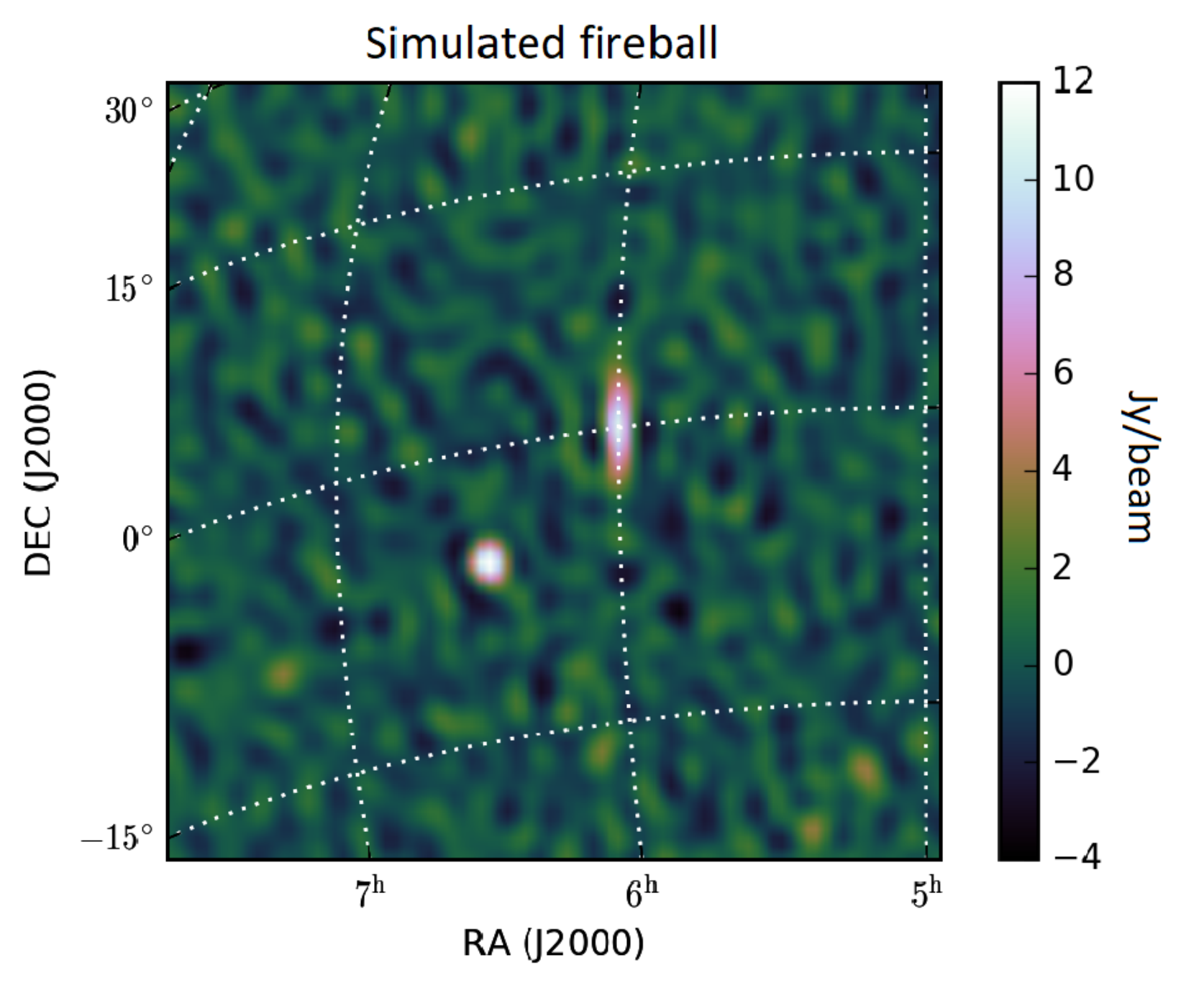}
	\caption{Simulated meteor on a difference image. The spectral index was set to -4, the length set to 4\degr as a point source observed by the LWA1, while the width of meteor set to 2\arcsec. The meteor trail emission was simulated with Gaussian distribution.}
	\label{fig:simulation}
\end{figure}

\begin{figure}
	\includegraphics[width=\columnwidth]{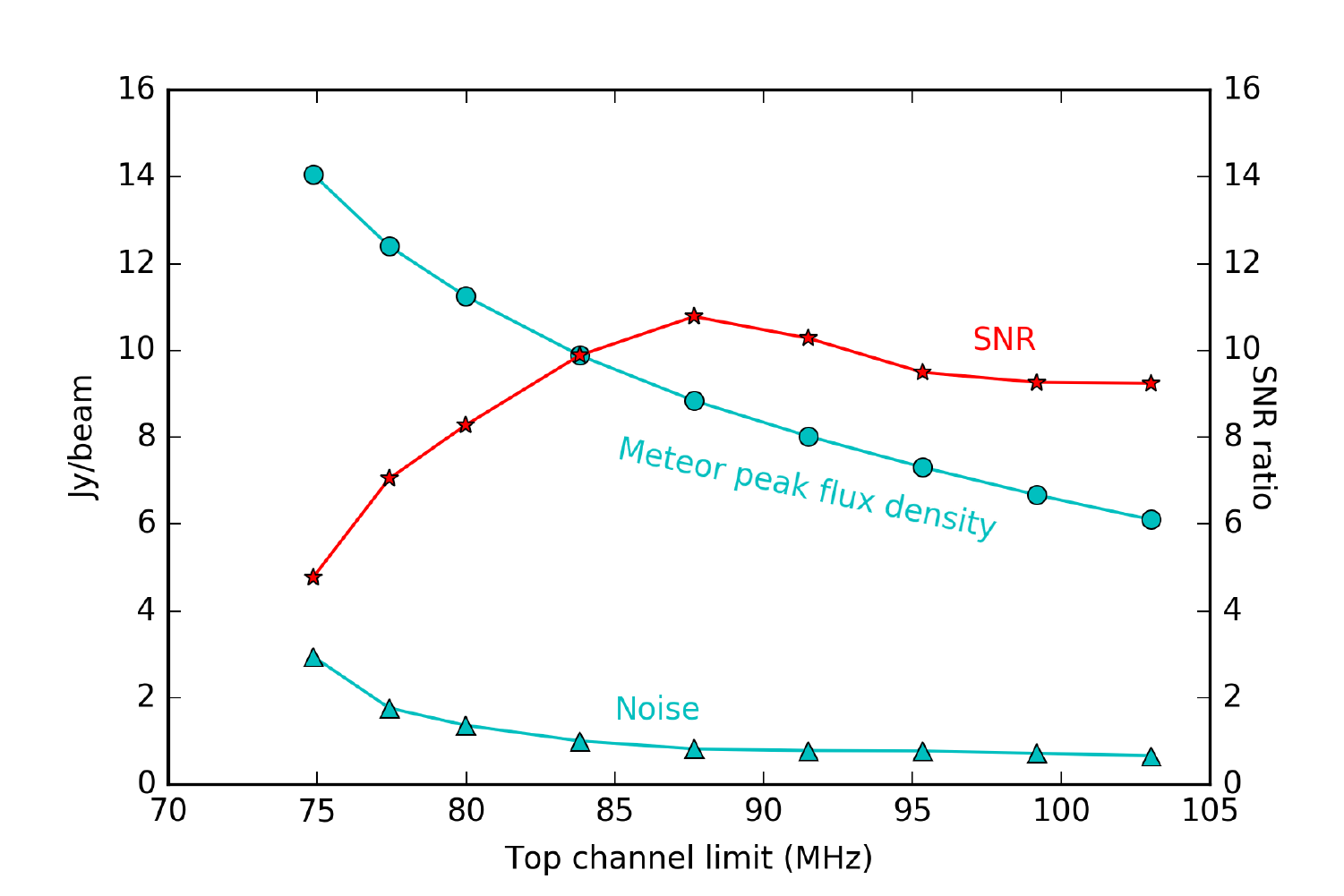}
	\caption{Relation between top channel limit and SNR for the simulated meteor when baseline limit was set to 120 m. The bottom channel is 72.3 MHz in all the tests.}
	\label{fig:channel}
\end{figure}

\begin{figure}
	\includegraphics[width=\columnwidth]{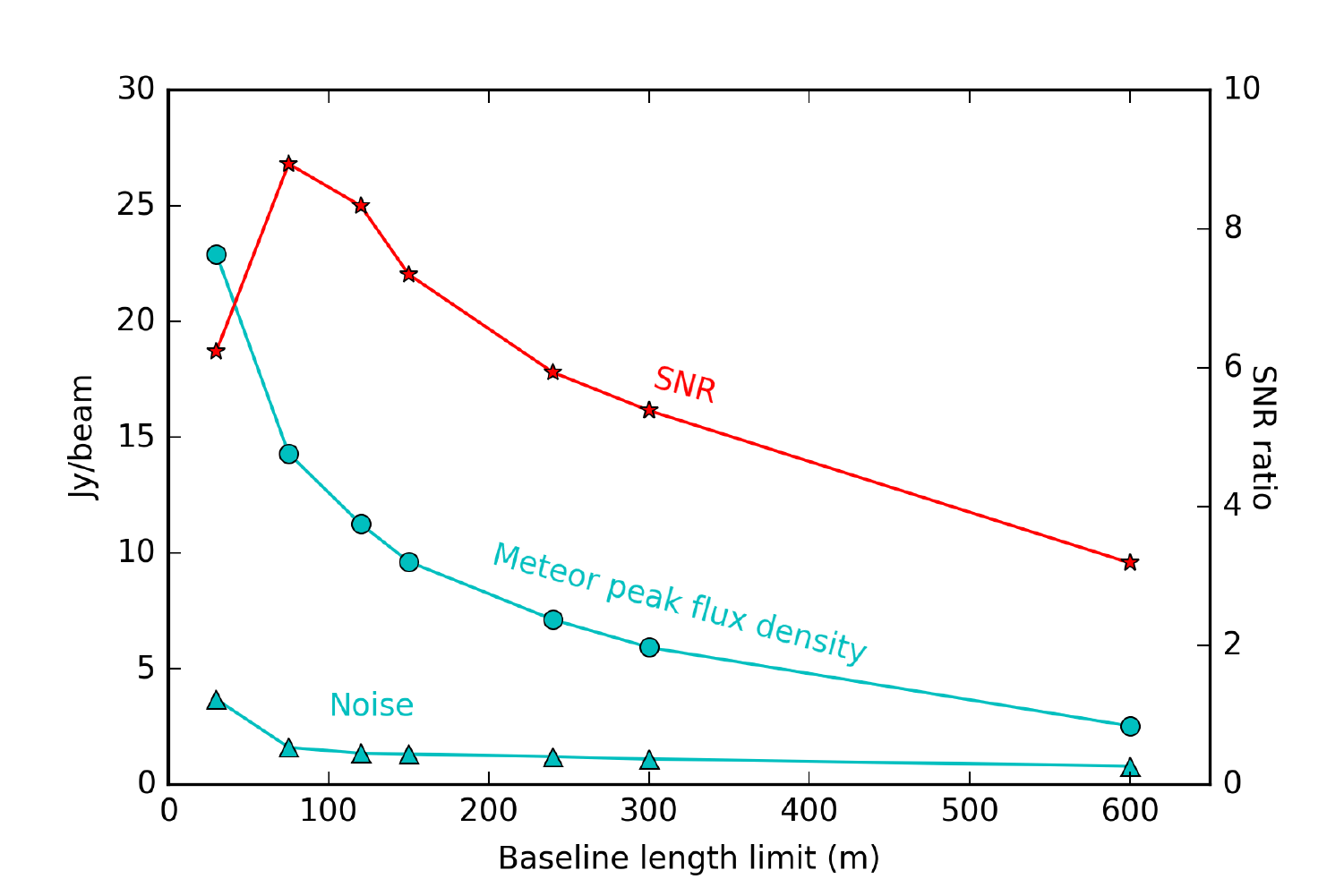}
	\caption{Relation between baseline length limit and SNR for the simulated meteor when top channel was set to 80 MHz.}
	\label{fig:baseline}
\end{figure}

We used CASA to put the simulated meteor into the visibility data, and made difference emission images with it (See Figure \ref{fig:simulation}). The spectral index adopted was -4. To get the best detectability, we made images with baseline upper limits from 30 m to 600 m, and top channel limits from 74.9 MHz to 103.2 MHz. The relation between the signal-to-noise ratio (SNR) and the limits are given in Figure \ref{fig:channel} and Figure \ref{fig:baseline}. According to these figures, the SNR reaches a peak near channel range 72.3 - 87.7 MHz and baseline length upper limit 75 m. However, considering other factors like overlapping with the FM band, sensitivity and spatial resolution, the channel range was set to 72.3 - 86.4 MHz, and the baseline length upper limit was set to 120 m (or 32 $\lambda$ at 80 MHz).

\subsection{Source finding}
The source finding is done by Aegean \citep{hancock2012compact, 2018PASA...35...11H}, which is designed to detect and characterize sources within radio images; it works by grouping pixels above a given threshold into contiguous groups called islands. Aegean also includes a Background And Noise Estimator (BANE) which provides a method for creating background and noise images.
In our data processing pipeline, we ran both BANE and Aegean on the difference emission images to extract meteors. BANE first formed detailed background maps for the images, then Aegean searched for pixels above the 5$\sigma$ level and grouped them together with nearby pixels above 3$\sigma$ level into ``islands". The outputs from Aegean included sizes of the islands and their integrated flux densities.


Aegean found approximately $2 \times 10^5$ islands in $1.2 \times 10^5$ images, so we did some selection based on the sizes of meteor events observed by the LWA1 and the projection method. Islands with a maximum angular size larger than 5\degr and pixel number over 10 were selected to form a list of meteor candidates. The process was able to detect our simulated meteor.





\section{Results}
\label{sec:results}

Using the data reduction pipeline described above, 5372 events were selected as meteor candidates in our 322-hour survey. However, most of them can be attributed to variations in bright radio sources caused by instrumental and/or ionospheric effects. No candidate was confirmed to be a meteor.

We followed a three-step method to check if a candidate was a meteor. First, we compared the candidate with its corresponding normal emission image. If the candidate was coincident with a bright radio source in the normal emission image (like the Fornax A event in Figure \ref{fig:frame4}), we believe that the candidate was related to the bright radio source, i.e. not a meteor. In this way, we excluded the majority of our candidates. Second, candidates not related to bright sources in normal images were compared with a subset (above 10 Jy) of the GLEAM catalogue. If a candidate was coincident with a radio source in the catalogue, the candidate would be excluded. Third, for the few candidates which could not be attributed to variabilities in bright radio sources, we checked corresponding optical images from DFN and the reflection images. If a candidate was consistent with an optical meteor or a reflection event, both spatially and temporally, it would be considered a probable event for intrinsic radio emission. However, none of the candidates were consistent with any optical meteors or reflection events.



\section{Discussion}
\label{sec:discussion}

When an experiment returns a null result, there are two possibilities: (1) no events were observed because no events occurred; (2) events occurred, but noise or timing prevented detection. In other words, the null result can be attributed to event rate density, flux density, or duration.

Here we provide an analysis of the sensitivity limits of our observational data and interpret the null result in terms of the physical parameters of intrinsic emission from meteors.
The analysis is based on a framework by \citet{trott2013framework}, which was designed to determine constraints on the detection rate of fast transient events. This framework takes into account the primary beam shape, frequency effects, and detection efficiency, resulting in the 2D probability distributions in the sensitivity-rate parameter space. 

\subsection{Minimum detectable flux density}
When meteor emission events do occur, detection is limited by the MWA beam pattern (Sokolowski et al, in prep) and sampling timescale. We take an unresolved meteor event for example.
If this event is detected, its signal $P_S$ must exceed a threshold, given by the noise $P_N$ and some SNR value, $C$. 
\begin{equation}
P_S>C P_N
\end{equation}

\begin{figure}
	\includegraphics[width=\columnwidth]{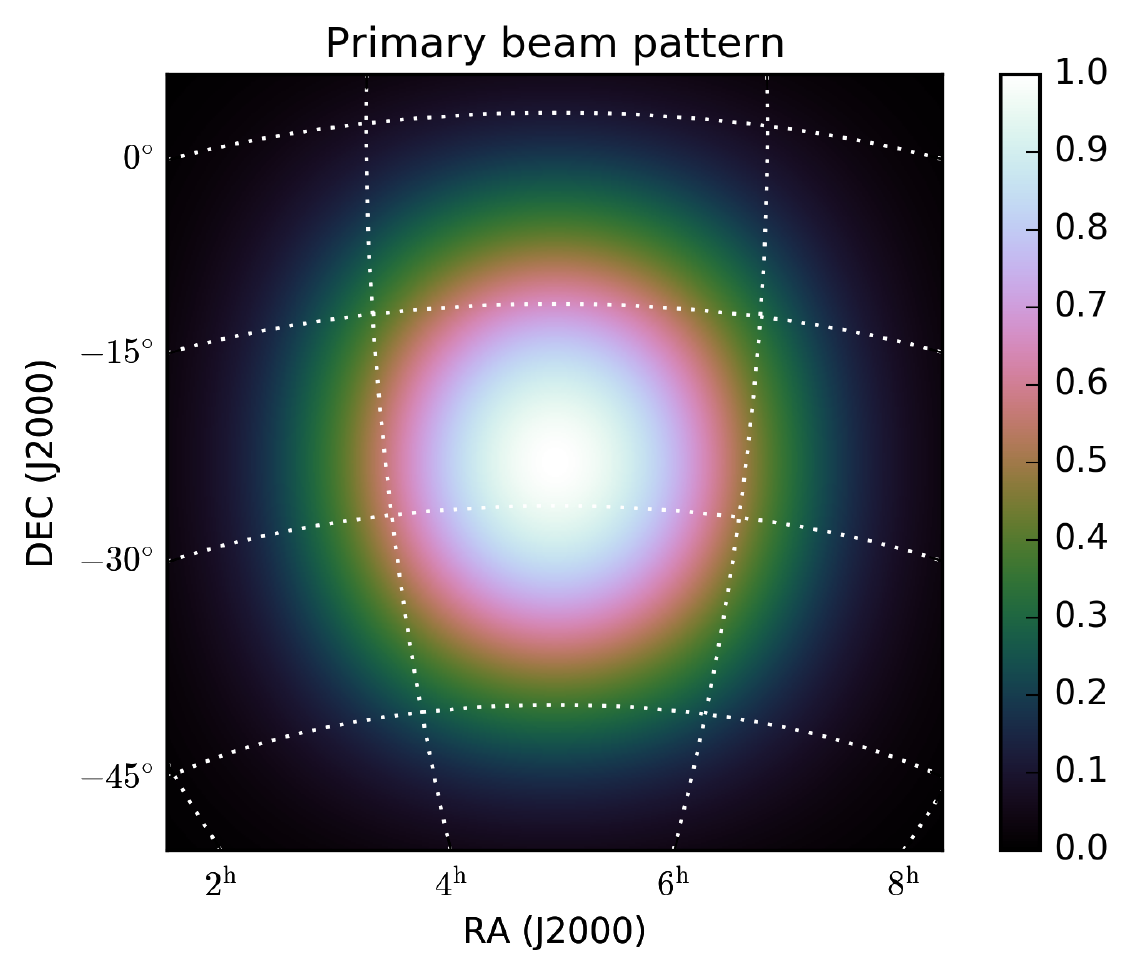}
	\caption{Beam model of the MWA at 80 MHz. Antennas pointed to zenith.}
	\label{fig:primary}
\end{figure}

When an array is used to detect radio signals, $P_S$ can be given as:
\begin{equation}
P_S=\frac{1}{\Delta \nu} \int_{\Delta \nu} B(\theta) P(\nu)  d\nu
\end{equation}
where $\Delta \nu$ is the bandwidth, $B(\theta)$ is the beam model (see Figure \ref{fig:primary}), $P_\nu$ is the frequency-dependent flux density of the radio source. We assume that the dependence of beam model on frequency can be neglected in our channel range. 
The noise $P_N$ can always be given from the System Equivalent Flux Density (SEFD):
\begin{equation}
P_N=\frac{\mathrm{SEFD}}{\sqrt{n_\mathrm{p} N_\mathrm{ant} (N_\mathrm{ant}-1) \Delta \nu \Delta t_s}}
\end{equation}
where $n_\mathrm{p}$ is the number of polarizations, $N_\mathrm{ant}$ is the number of antennas within the MWA, $\Delta t_s$ is the sampling timescale.

It is assumed that the flux density of a meteor can be represented by a power law in frequency, so we have $P(\nu)=S_0 (\nu/\nu_0)^\alpha$, where $S_0$ is the flux density at the reference frequency, $\nu_0$. 



For some short-duration meteor events, the temporal sampling time may exceed the duration of meteor emission, thus the radio signal received suffers a loss in flux density due to the averaging of the signal over time. Here we introduce the Duration Threshold factor $\eta$, which is defined as
\begin{equation}
\eta=\mathrm{min}(\Delta t_\mathrm{act}/\Delta t_s,1)
\end{equation}
where $\Delta t_\mathrm{act}$ is the duration of meteor emission, $\Delta t_s$ is the sampling time. Then we have
\begin{equation}
S_\mathrm{measure} \approx \eta S_\mathrm{act} =
\begin{cases}
S_\mathrm{act}, \ \ \quad \quad \mathrm{for}\quad \Delta t_\mathrm{act} \geqslant \Delta t_\mathrm{s} \\
\frac{\Delta t_\mathrm{act}}{\Delta t_\mathrm{s}}S_\mathrm{act} , \quad \mathrm{for} \quad \Delta t_\mathrm{act} < \Delta t_\mathrm{s}
\end{cases}
\end{equation}
where $S_\mathrm{measure}$ is the measured signal flux density, $S_\mathrm{act}$ is the original signal flux density. 


Thus the minimum detectable original flux density at angle $\theta$ from beam centre, $S_\mathrm{min} (\theta)$, can be given by:
\begin{equation}
S_{\mathrm{min}} (\theta) = \frac{C}{\eta}~P_N~(\frac{1}{\Delta \nu} \int_{\Delta \nu} B(\theta) (\nu/\nu_0)^\alpha  d\nu)^{-1}
\end{equation}

\begin{figure}
	\includegraphics[width=\columnwidth]{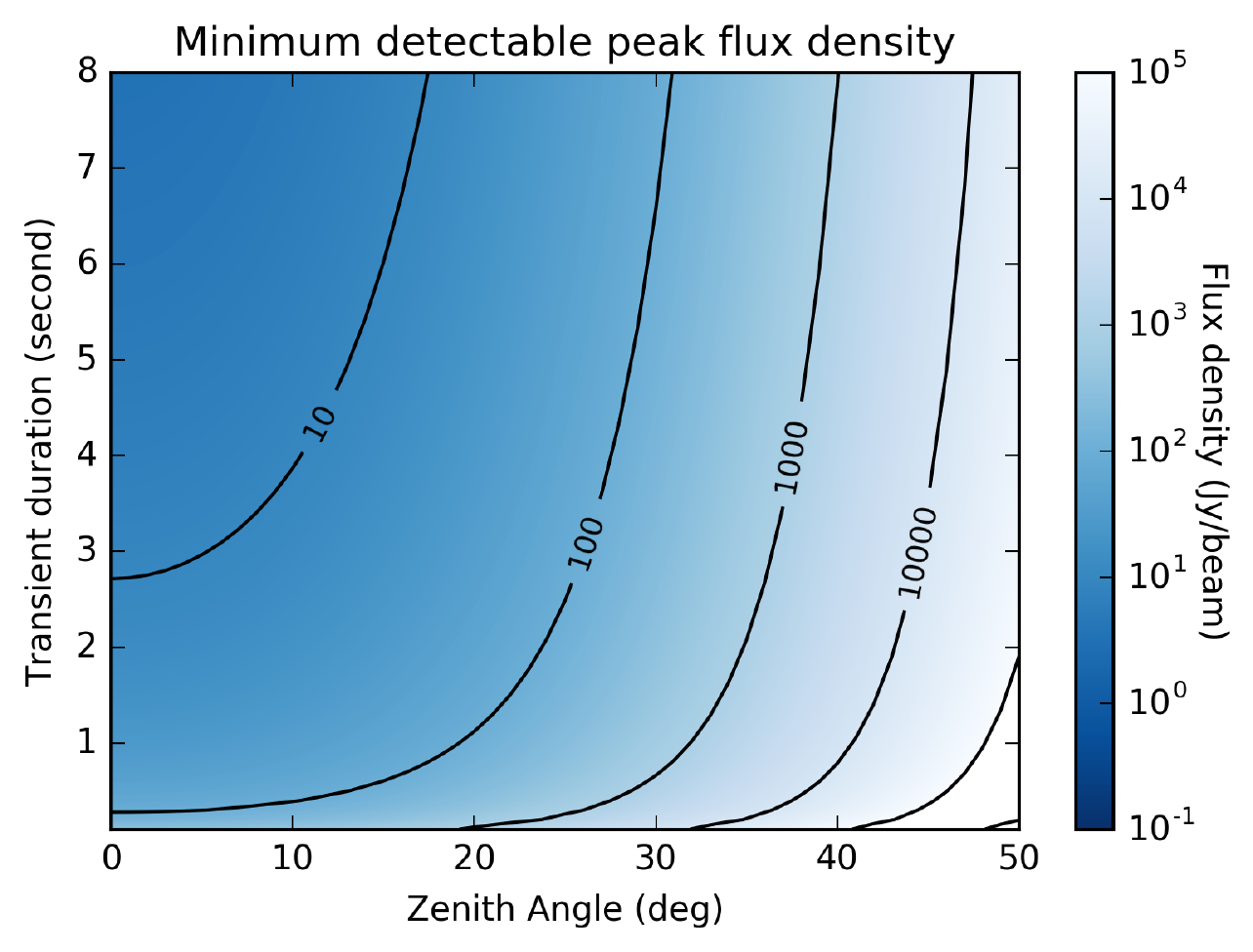}
	\caption{Minimum peak flux density required for meteor event detection within our observations. Detection is limited by the MWA beam model and the meteor event duration. The sampling timescale is 8 seconds. Events with durations longer than 8 seconds have the same minimum detectable peak flux density as 8-second events.}
	\label{fig:detection}
\end{figure}

Emission from meteors can be resolved as observed by the MWA, so detection is limited by the peak flux density. We estimated the minimum detectable peak flux density in our observations, which is illustrated in Figure \ref{fig:detection}.

\subsection{Probability of detection}

However, the detected flux density of a radio source is composed of its true flux density $S$ and noise. The noise follows a Gaussian distribution, with mean value $\mu = 0$ and variance $\sigma^2$. Thus the probability for a radio source with true flux density $S$ to be detected above the threshold, $C\sigma$, is given by the Cumulative Distribution Function (CDF):
\begin{equation}
P(S>C\sigma)=\int^\infty_{C\sigma}\mathcal{N}(S,\sigma^2)dS
=\frac{1}{2}+\frac{1}{2}\mathrm{erf}(\frac{S-C\sigma}{\sqrt{2}\sigma})
\end{equation}
where $\mathcal{N}(S,\sigma^2)$ denotes the Gaussian distribution and erf is the error function. The probability that an event is not detected because of noise is the complementary function, $1-P(S>C\sigma)$.

We assume that meteor afterglows are randomly distributed (both temporally and spatially) with a mean frequency of occurrence. Then the probability that $k$ events occur with an expectation of $\lambda$ follows the Poisson distribution:
\begin{equation}
P(k;\lambda)=e^{-\lambda}\frac{\lambda^k}{k!}
\end{equation}

The probability that at least one event should be detected is:
\begin{equation}
P=1-P(0;\lambda)=1-e^{-\lambda}
\end{equation}

\begin{figure}
	\includegraphics[width=\columnwidth]{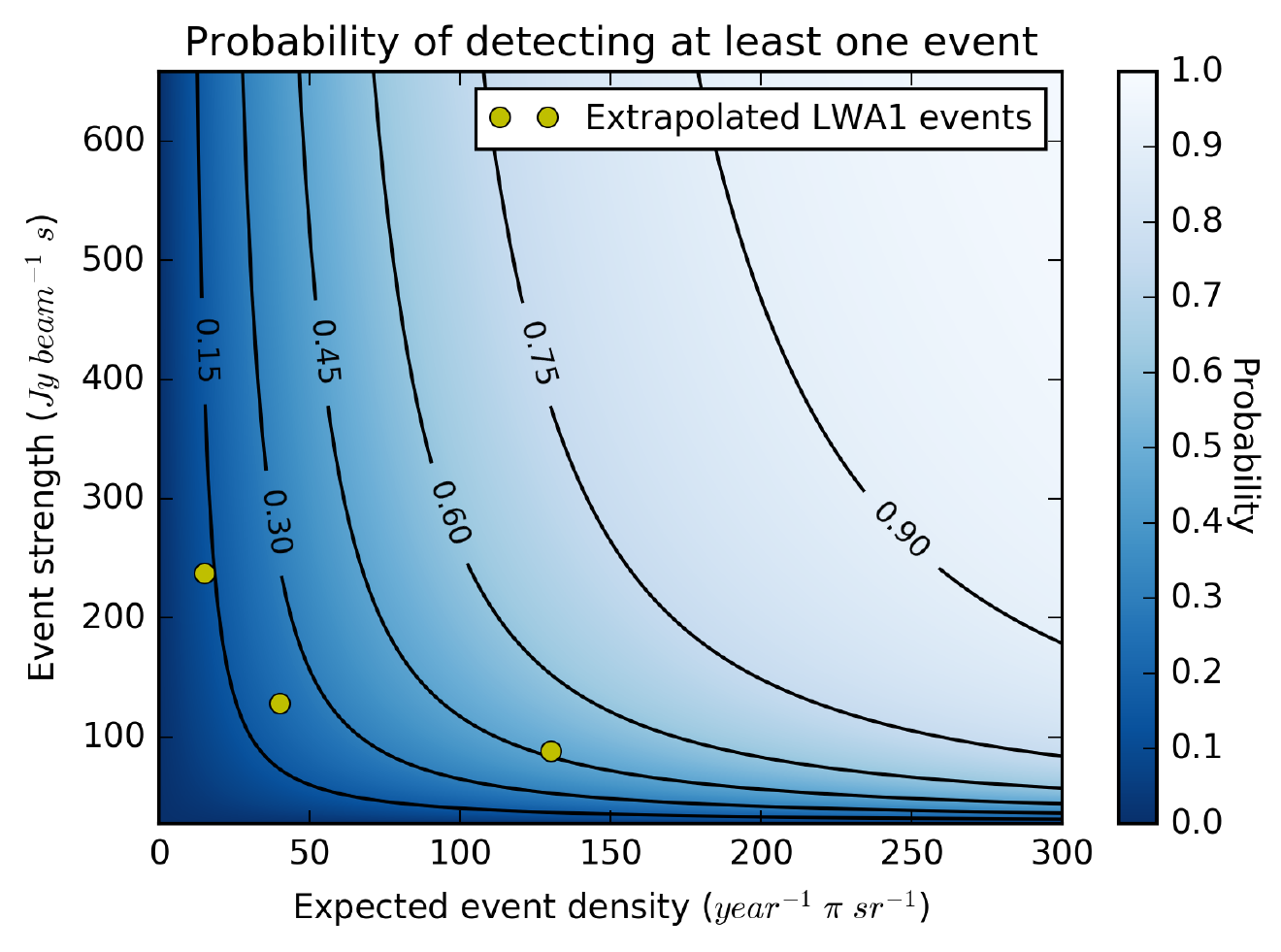}
	\caption{Probability of detecting at least one meteor in our observation. Yellow circles indicate the extrapolated event strength (Equation \ref{eqn:strength}) of three LWA1 events \citep{obenberger2016rates} at MWA frequencies, assuming a spectral index of -4. Durations of the three LWA1 events are about 40 seconds at LWA1 band.}
	\label{fig:rate}
\end{figure}

Figure \ref{fig:rate} shows the probability of detecting at least one meteor in our observations, given expected event density and event strength. The ``event strength" is defined as 
\begin{equation}
\label{eqn:strength}
S_\mathrm{act} \eta \Delta t_\mathrm{s} =
\begin{cases}
S_\mathrm{act} \Delta t_\mathrm{s}, \ \ \quad \mathrm{for}\quad \Delta t_\mathrm{act} \geqslant \Delta t_\mathrm{s} \\
S_\mathrm{act} \Delta t_\mathrm{act}, \quad \mathrm{for} \quad \Delta t_\mathrm{act} < \Delta t_\mathrm{s}
\end{cases}
\end{equation}
to cover events both shorter and longer than the sampling time. The total effective observational time is 308 hours, and the frequency range is 72-86 MHz. The rise of meteor event numbers in our observations caused by meteor showers is included. 

In Figure \ref{fig:rate} we indicate three LWA1 events with event densities 15, 40 and 130 times $year^{-1}~\pi~sr^{-1}$, as described in \citet{obenberger2016rates}. A spectral index of -4 is used to extrapolate flux densities of these events to the MWA frequency range. As illustrated in Figure \ref{fig:rate}, the probabilities for the MWA to detect these three events in our observations are less than 50\%.

\subsection{An upper limit on meteor spectral index}
According to \citet{obenberger2016rates}, the luminosity function of meteor emission resembles a power law, with higher event rates for fainter meteors. Under the assumption of a spectral index of -4, the three LWA1 events would be above $10\sigma$ at MWA frequencies.  Under this assumption, fainter meteors with higher event rates above the $5\sigma$ detection threshold of the MWA should be abundant.
Given we make no such 5$\sigma$ detections, we give an estimated relation between meteor flux density and event rates, based on which we can derive an upper limit on the spectral index of meteor emission, relevant for our frequency range.

We start with the radio magnitude of meteors, which is defined by the ionization produced per unit length of the meteor path (without any reference to visual luminosity).
An approximate radio magnitude relation is deduced by \citet{mckinley1961meteor}:
\begin{equation}
M_r=40-2.5 \log_{10} q
\end{equation}
where $M_r$ is the radio magnitude of a meteor, $q$ is the electron line density.

According to previous radio observations by radar, there is an empirical relation between meteor radio magnitude and meteor numbers \citep{mckinley1961meteor}:
\begin{equation}
\log_{10}N(<M_r) \simeq 16 - 1.34 \log_{10} q
\end{equation}
where $N(<M_r)$ is the total number of meteors of radio magnitude $M_r$ and brighter encountered by the Earth's atmosphere in a 24-hour period.

The mechanism for intrinsic radio emission from meteors is still under discussion, with one possible explanation being radiation of Langmuir waves \citep{obenberger2015dynamic}. Thus we assume that the peak flux density of meteor emission is proportional to the peak electron line density, i.e. $S \propto q$ (the width of ionized meteor trails can be neglected due to the low spatial resolution of the MWA images made in this project). By differentiating the cumulative number of meteors we have
\begin{equation}
\label{eqa}
\mathrm{d} N_S \propto S^{-2.34}
\end{equation}
where $\mathrm{d} N_S$ is the number of meteors between peak flux density $S$ and $S+\mathrm{d}S$. Since all the meteor events observed by the LWA1 have durations longer than our sampling time, and significant drops in meteor spectral indices occur after the first 20 seconds for the three typical events, we estimate that the durations for most meteors detectable to the MWA also exceed the sampling time, i.e. $\eta = 1$. Thus meteor event strength is proportional to peak flux density in our project.

\begin{figure}
	\includegraphics[width=\columnwidth]{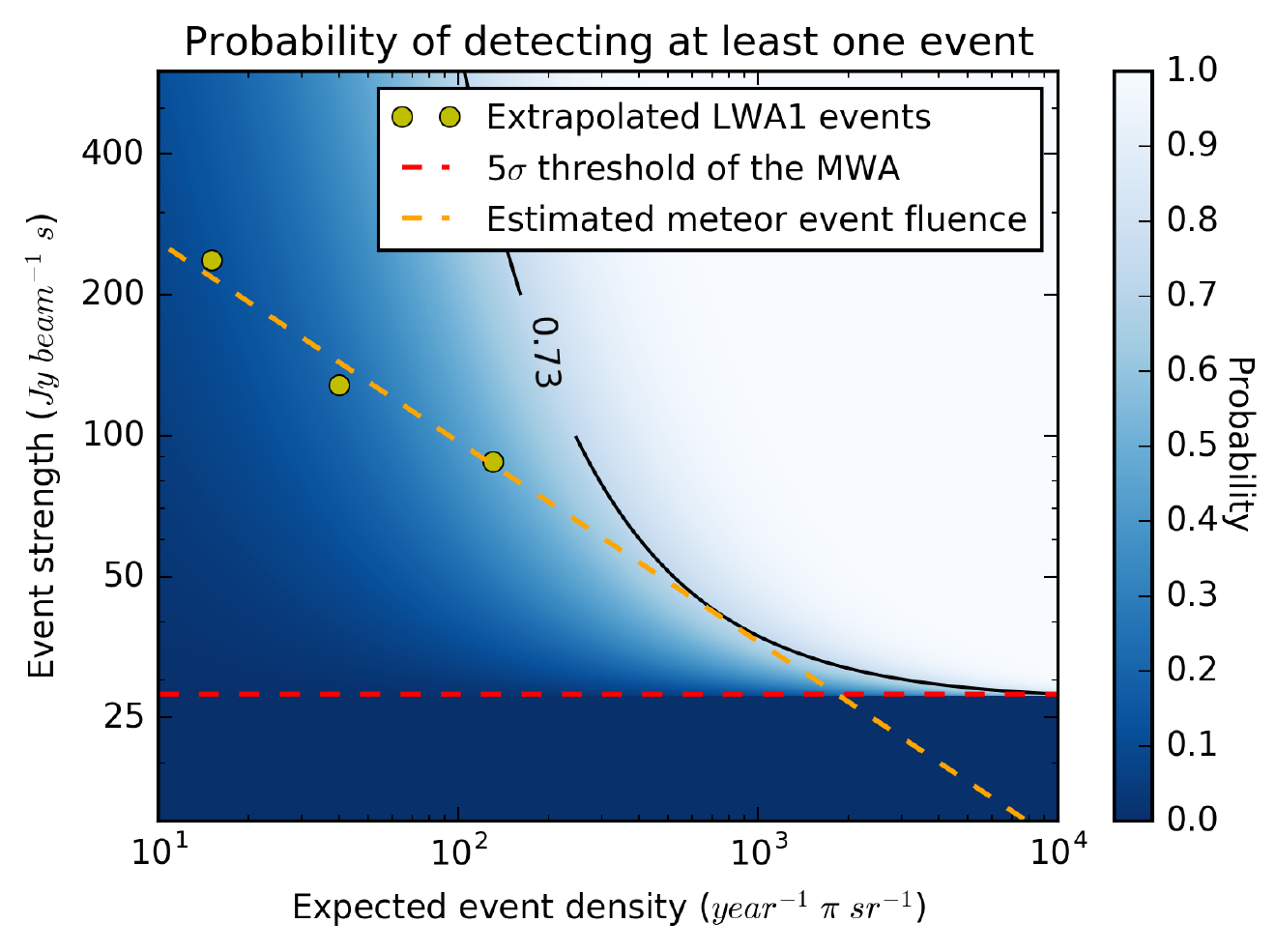}
	\caption{Estimated luminosity function of intrinsic radio emission from meteors. Yellow circles are the extrapolated LWA1 events \citep{obenberger2016rates} at MWA frequencies, under an assumed spectral index of -4; dashed orange line represents the estimated relation between meteor event strength and event density, as given in Equation \ref{eqa}; dashed red line represents the $5\sigma$ threshold of the MWA.}
	\label{fig:MWA}
\end{figure}

Based on the three meteor events and Equation \ref{eqa}, we give an estimated relation between event strength and expected event density, as illustrated in Figure \ref{fig:MWA}. 
It is shown that in our survey, the probability of detecting a fainter event is higher than that of the three LWA1 events, but not exceeding 73\%.

\begin{figure}
	\includegraphics[width=\columnwidth]{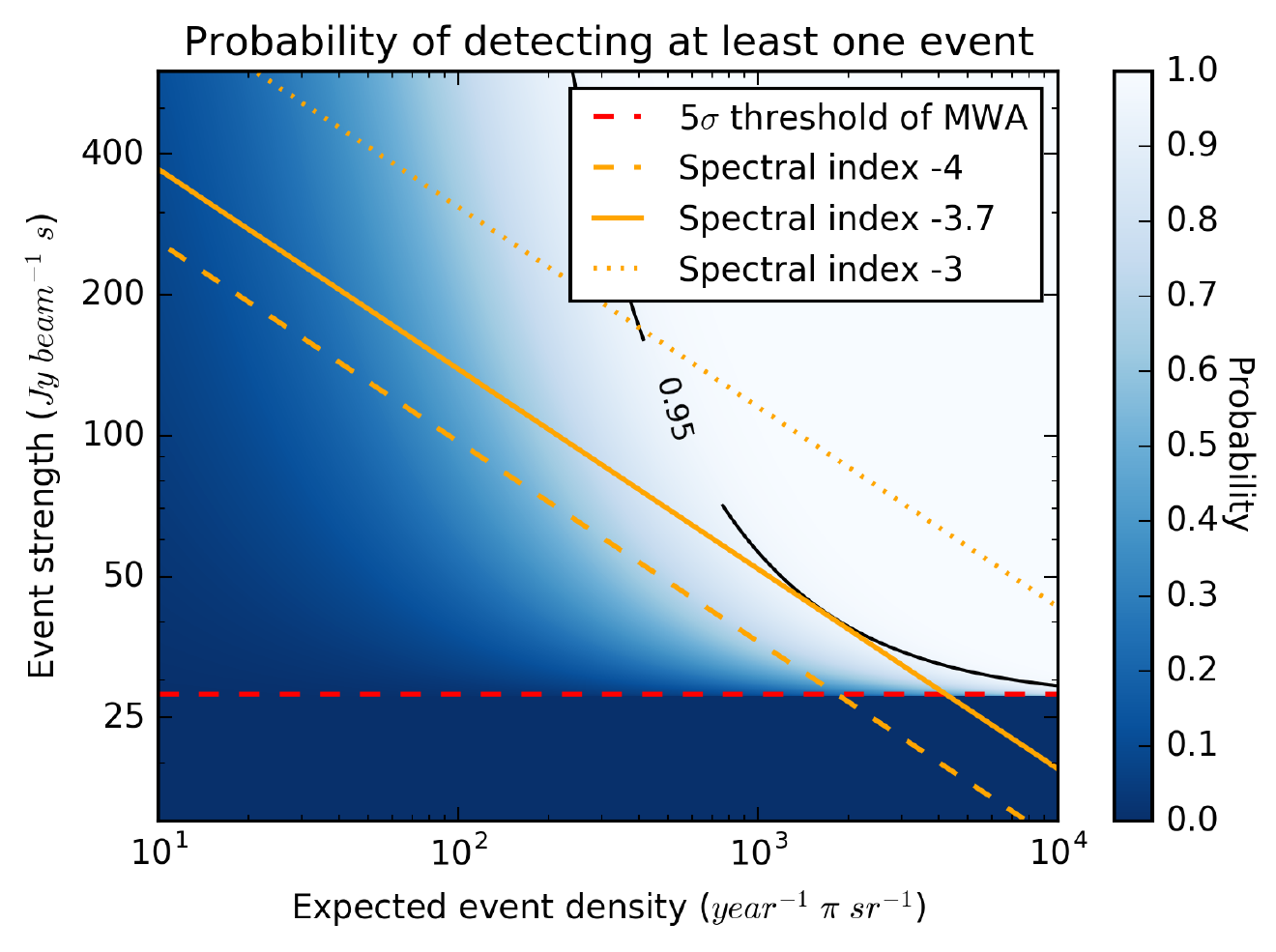}
	\caption{Luminosity function of intrinsic radio emission from meteors, derived with different spectral indices. The dotted, solid and dashed orange lines give estimated meteor event strength with spectral indices -3, -3.7 and -4. }
	\label{fig:index}
\end{figure}

However, the extrapolated LWA1 events and the estimated meteor strength-rate relation given in Figure \ref{fig:MWA} are derived using spectral index -4. If we use spectral index -3, then the probability for the MWA to detect faint meteors in our survey exceeds 95\% (see Figure \ref{fig:index}). At spectral index -3.7, the probability of detecting at least one event in our observations is 95\%. In other words, we give an estimated meteor spectral index upper limit of -3.7 with 95\% confidence.

\subsection{Other transient events}

Besides meteor reflections and scintillations, some other transient events were also captured in the FM band. Previously, \citet{mckinley2012low} and \citet{tingay2013detection} detected reflected FM signals from the Moon and the International Space Station, respectively, with the MWA.

Two of the detected transient events found in our data lasted for minutes and moved tens of degrees across the sky. We overplotted radio images with the positions of satellites and confirmed that these two events were caused by two satellites, Duchifat-1 and Alouette-2. The orbital parameters\footnote{\url{https://www.space-track.org/}} of the satellites are listed in Table \ref{tab:sate}. Since neither of the two satellites transmits in the observed FM band, we believe that the transient events caused by the satellites were due to radio reflection of terrestrial FM broadcasting signals. A detailed discussion of the satellite reflection events will be published in the future (Hancock et al, in prep).

\begin{table}
	\centering
	\caption{Parameters of the two satellites observed by the MWA. RCS is short for Radar Cross Section.}
	\label{tab:sate}
	\begin{tabular}{ccc} 
		\hline
		Parameters  & Duchifat-1 & Alouette-2 \\
		\hline
		Period (min) & 96.65 & 117.52 \\
		Inclination (deg) & 97.91 & 79.80 \\
		Apogee (km) & 608 & 2637 \\
		Perigee (km) & 588 & 502 \\
		RCS size range ($\mathrm{m^2}$) & <0.1 & 0.1<RCS<1.0 \\
		\hline
	\end{tabular}
\end{table}

\subsection{Potential for other facilities}

Considering the spectral index and event rate of meteor emission, the main factors that limit a radio telescope's ability to detect radio emission from meteors are frequency range, Field-of-View (FoV) and sensitivity.

Here we give the parameters of some low frequency arrays in Table \ref{tab:compare}. The MWA has a high sensitivity, but its meteor detection is restricted by the relatively small FoV and the high frequency range. The LWA1 has two modes: PASI provides all-sky images, but the bandwidth is narrow, while the phased array mode has a wide bandwidth with a very small FoV. With these two modes, the LWA1 has detected more than 100 meteor emission events, but only obtained a few spectra. AARTFAAC, with its all-sky FoV, the suitable frequency range, and a wide bandwidth, has the best opportunity to collect radio spectra from meteor emission. Since the frequency range of AARTFAAC overlaps with both the LWA1 and the MWA (lower), it would be possible to directly determine the higher frequency behaviour of events detected by the LWA1.

\begin{table*}
	\centering
	\caption{Parameters of contemporary radio transient detection arrays. The parameters are estimated at 80 MHz for the MWA, 60 MHz for AARTFAAC, and 74 MHz for the LWA1 \citep{wijnholds2011situ, tingay2013murchison, ellingson2013lwa1, prasad2014real}. Only parameters of the low band antennas (LBA) of AARTFAAC are listed here. }
	\label{tab:compare}
	\begin{tabular}{ccccc} 
		\hline
		Parameter  & MWA & LOFAR (AARTFAAC) & LWA1 (PASI) & LWA1 (phased array mode) \\
		\hline
		Frequency range (MHz) & 80-300 & 30-80 & 10-88 & 10-88 \\
		Field of view (sr) & 0.06$\pi$ & $\pi$ & $\pi$ & 0.005$\pi$ \\
	    Total effective area (m$^2$) & 3016 & 2617 & 1393 & 1393 \\
	    $T_\mathrm{sys}(\nu^{-2.55} \mathrm{K})$ & 1730 & 3600 & 2100 & 2100 \\
	    Angular resolution (arcmin) & 3 & 60 & 120 & 120 \\
	    Spectral resolution (kHz) & 10 & 16 & 75 & 19.14 \\
	    Bandwidth (MHz) & 30.72 & 13 & 0.075 & 36 \\
	    Temporal resolution (s) & 0.5 & 1 & 5 & 0.04 \\
		\hline
	\end{tabular}
\end{table*}

\section{Conclusions}
\label{sec:conclusions}

We have reported a survey for intrinsic radio emission from meteors with the MWA, which spans observing frequencies from 72.3 to 103.0 MHz. Optical observations were also carried out to verify possible candidates. Although radio reflection from ionized meteor trails and optical meteors were detected in the survey, no intrinsic emission was observed. Assuming the radio emission from meteors follows a power law with frequency, we derived an upper limit -3.7 on meteor emission spectral index with a confidence of 95\%. This upper limit is consistent with the previous estimation from \citep{obenberger2016rates}.

We have also reported the detections of some other transient events, including the reflected FM broadcast signals from satellites in low Earth orbits, which are consistent with previous simulations by \citet{tingay2013detection}. These detections show the potential of the MWA for Space Situational Awareness (SSA).

\section*{Acknowledgements}

This scientific work makes use of the Murchison Radio-astronomy Observatory, operated by CSIRO. We acknowledge the Wajarri Yamatji people as the traditional owners of the Observatory site. Support for the operation of the MWA is provided by the Australian Government (NCRIS), under a contract to Curtin University administered by Astronomy Australia Limited. We acknowledge the Pawsey Supercomputing Centre which is supported by the Western Australian and Australian Governments. Parts of this work are supported by the Australian Research Council Centre of Excellence for All-Sky Astrophysics (CAASTRO), funded through grant number CE110001020, and the Australian Research Council Centre of Excellence for All Sky Astrophysics in 3 Dimensions (ASTRO 3D), funded through grant number CE170100013.

We acknowledge the financial support from China Scholarship Council (Grant No. 201504910639). This work is also supported by the National Natural Science Foundation of China (Grant Nos. 11473073, 11573075, 11661161013, 11633009), CAS Interdisciplinary Innovation Team, and Foundation of Minor Planets of the Purple Mountain Observatory. 




\bibliographystyle{mnras}
\bibliography{meteor} 








\bsp	
\label{lastpage}
\end{document}